\documentclass{webofc}
\usepackage[varg]{txfonts}   
\usepackage{comment}
\usepackage{braket}
\usepackage{color}

\begin{document}
\title{ New calculation of collision integrals for cosmological phase transitions}
%
%

\author{
		\firstname{Carlo} \lastname{Branchina}\inst{1,2} \fnsep\thanks{\email{carlo.branchina@unical.it}},
		\firstname{Angela} \lastname{Conaci}\inst{1,2} \fnsep\thanks{\email{angela.conaci@unical.it}}
		\firstname{Stefania} \lastname{De Curtis}\inst{3}\fnsep\thanks{\email{stefania.decurtis@fi.infn.it}} \and
        \firstname{Luigi} \lastname{Delle Rose}\inst{1,2}\fnsep\thanks{\email{luigi.dellerose@unical.it}} \and
        \firstname{Andrea} \lastname{Guiggiani}\inst{3}\fnsep\thanks{\email{andrea.guiggiani@unifi.it}} \and
        \firstname{\'Angel} \lastname{Gil Muyor}\inst{4}\fnsep\thanks{\email{agil@ifae.es}} \and
        \firstname{Giuliano} \lastname{Panico}\inst{3}\fnsep\thanks{\email{giuliano.panico@unifi.it}}
}

\institute{
Dipartimento di Fisica, Universit\`a della Calabria, I-87036 Arcavacata di Rende, Cosenza, Italy
\and
 INFN, Gruppo Collegato di Cosenza, Arcavacata di Rende, I-87036, Cosenza, Italy
\and
INFN Sezione di Firenze and Dipartimento di Fisica e Astronomia, Universit\`a di Firenze, Via G. Sansone 1, I-50019 Sesto Fiorentino, Italy
\and
IFAE and BIST, Universitat Aut\`onoma de Barcelona, 08193~Bellaterra,~Barcelona,~Spain
          }

\abstract{First order phase transitions in the early universe may have left a variety of experimentally accessible imprints. The dynamics of such transitions is governed by the density perturbations caused by the propagation of the bubble wall in the false vacuum plasma, conveniently described by a Boltzmann equation. The determination of the bubble wall expansion velocity is crucial to determine the experimental signatures of the transition. We report on the first full (numerical) solution to the Boltzmann equation. Differently from traditional ones, our approach does not rely on any ansatz. The results significantly differ from the ones obtained within the fluid approximation and large differences for the friction acting on the bubble wall are found. The wall velocity is calculated in a singlet extension of the Standard Model, including out-of-equilibrium contributions from both the top quark and the electroweak gauge bosons.}
\maketitle

\section{Introduction}\label{sec:intro}

The recent observation of gravitational waves has fueled a lively interest in the study of cosmological phase transitions (PhTs) and their dynamics. If a first order phase transition towards the electroweak (EW) vacuum did occur in the early universe, the latter is expected to have produced a source of gravitational waves that could be detected by future space-based interferometers \cite{Caprini:2015zlo,Caprini:2019egz,Kawamura:2006up,Kawamura:2011zz,Hu:2017mde,Ruan:2018tsw,TianQin:2015yph}. This would make theories of the EW phase transition (EWPT) falsifiable and, combined with collider experiments, could play a key role in our quest for beyond the Standard Model (BSM) physics. As it is well known, in fact, the Standard Model cannot account for a first order EWPT, that can only be realized in BSM scenarios. 

Besides the gravitational wave background, a first order PhT would have left many other imprints, and could for instance be responsible for the abundance of dark matter, the matter-antimatter asymmetry, and many other phenomena.
In this respect, given a BSM model and the transition pattern from the high temperature vacuum to the EW one, an accurate modelling of the PhT dynamics is crucial to quantitatively assess its impact.
The transition is induced by a non-trivial scalar profile that describes a spherically symmetric solution \cite{Coleman:1977py,Callan:1977pt,Linde:1977mm} where the scalar field is close to its true vacuum value in a region of space and tends to its false vacuum one at larger distances. When the energy difference between the two vacua is sufficiently small, the scalar field varies rapidly in a small region, and the solution can be thought of as describing a bubble of true vacuum separated by a wall from a false vacuum background. 
The transition then proceeds by nucleation and expansion of these bubbles.

One of the fundamental quantities that characterizes the gravitational wave spectrum, as well as other quantities of interest, is the expansion speed of the PhT front, typically referred to as the bubble wall (terminal) velocity $v_w$. 
In the steady state regime, the latter arises from the competition of two different contributions: the external friction exerted by particles in the false vacuum sea (modelled as a plasma) hitting the wall, and the internal pressure caused by the potential difference between the false and true vacuum phases. The expanding bubble drives the plasma out of equilibrium, inducing a backreaction that slows down its propagation. This is not the only source of friction. In fact, the non-trivial spatial profile of the plasma temperature and velocity hinders the acceleration of the bubble, irrespective of the species in the plasma being out of equilibrium or not.

The wall velocity at equilibrium was first studied in \cite{Ignatius:1993qn} and has recently re-attracted a large interest. Comparing the results obtained assuming local thermal equilibrium to those found including out-of-equilibrium effects from the top quark for some benchmark points in a singlet extension of the Standard Model (SSM), Ref.\,\cite{Laurent:2022jrs} advances that calculations of the wall velocity at equilibrium may provide a good approximation to the full result. Although it was later shown by some of the present authors that this expectation is not really fulfilled~\cite{DeCurtis:2023hil,DeCurtis:2024hvh}, the equilibrium determination of $v_w$ certainly remains of interest, especially in connection to the iterative procedure that some of us presented in \cite{DeCurtis:2022hlx}, through which out-of-equilibrium contributions can be systematically quantified. The wall velocity at equilibrium represents the zeroth order approximation of this calculation, and can be used as a first estimate to scan the parameter space of BSM theories~\cite{in preparation}.  

As first shown in \cite{Moore:1995ua,Moore:1995si}, perturbations around equilibrium can be described by an effective Boltzmann equation. The authors developed an approach, dubbed here ``old formalism" (OF), where the perfect fluid ansatz is taken for distribution functions and the Boltzmann equation is solved by taking suitable (weighted) moments that recast it into a system of ordinary differential equations. 
The fluid approximation has two main drawbacks, namely the appearance of a singularity in the friction for sonic walls and an unphysical dependence on the choice of weights: different weights lead to different results for the friction, and hence for the out-of-equilibrium effects. It has been recently argued that the singularity at the speed of sound is an artefact of the approximations used to solve the Boltzmann equation~\cite{Cline:2020jre,Laurent:2020gpg}, leading the authors to conclude that no critical behaviour is related to a sonic bubble wall.

Although the ``new formalism" (NF) developed in \cite{Cline:2020jre,Laurent:2020gpg} and an extended fluid approximation~\cite{Dorsch:2021nje,Dorsch:2021ubz} successfully removed the unphysical singularity, both these strategies rely on an ansatz to solve the Boltzmann equation. The dependence of the results on the various ansatz urges to find a complete solution that does not rely on any assumption. In \cite{DeCurtis:2022hlx} some of us presented for the first time a numerical solution to the full Boltzmann equation. To deal with the collision integral that appears in the equation, technically its most challenging part, an iterative procedure was introduced. The efficiency of the numerical algorithm was significantly improved in two subsequent works \cite{DeCurtis:2023hil,DeCurtis:2024hvh}, where it was observed that the symmetries of the collision integral allow it to be interpreted as an Hermitian operator. In turn, this makes it possible to study it through its spectral decomposition, where a hierarchy between eigenvalues was observed and exploited to simplify the calculation.

\section{The Boltzmann equation}\label{sec:Boltzmann}

Large bubbles can be conveniently described taking a planar approximation for the wall. Throughout this manuscript, we assume the dynamics has reached a steady state regime where the wall moves at a constant velocity $v_w$. Orienting the $z$-axis in the direction opposite to the propagation of the PhT front, in the wall reference frame the Boltzmann equation for the distribution function $f$ of a particle of mass $m(z)$ reads
\begin{equation}
{\cal L}[f] \equiv \left(\frac{p_z}{E} \partial_z - \frac{(m^2(z))'}{2E} \partial_{p_z}\right) f = - {\cal C}[f]\,,
\end{equation}
where $\mathcal L$ is the Liouville operator and  ${\cal C}$ is the collision integral that accounts for microscopic interactions between particles in the plasma. 

The collision integral ensures that at sufficiently large distances from the wall (local) thermal equilibrium is restored, so that for $z\to\pm\infty$ the distribution function tends to the Fermi or Bose--Einstein one,
\begin{equation}
f_v = \frac{1}{e^{\beta \gamma(E - v p_z)} \pm 1}\,,
\end{equation}
with $v$ the fluid velocity, $\beta = 1/T$ and $\gamma = (1- v^2)^{-1/2}$. Deviations from equilibrium are expected to be significant only close to the wall. When they are small enough, writing $f=f_v+\delta f$, the Boltzmann equation can be linearized in $\delta f$ to give
\begin{equation}\label{eq:Boltz_lin}
\left(\frac{p_z}{E} \partial_z - \frac{(m^2(z))'}{2E} \partial_{p_z}\right) \delta f + {\overline{\cal C}}[\delta f] = \frac{(m^2(z))'}{2E} \partial_{p_z} f_v = \beta \gamma v \frac{(m^2(z))'}{2E} f_v'\,,
\end{equation}
where we defined
\begin{equation}
f_v' \equiv - \frac{e^{\beta \gamma(E- v p_z)}}{(e^{\beta \gamma(E- v p_z)} \pm 1)^2}\,,
\end{equation}
and ${\overline{\cal C}}[\delta f]$ denotes the linearized collision integral.
The right hand side of \eqref{eq:Boltz_lin}, the source term, gives sizable contributions close to the wall and rapidly decays away from it. 

\begin{figure}
	\centering
	\includegraphics[width=.52\textwidth]{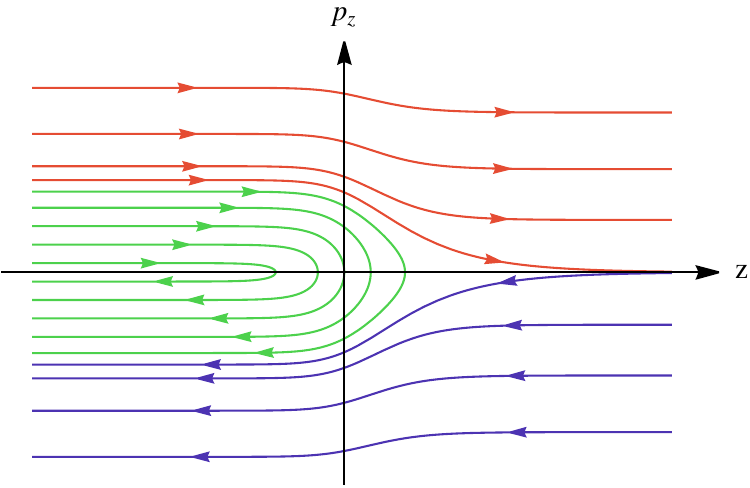}
	\hfill
	\raisebox{1em}{\includegraphics[width=.43\textwidth]{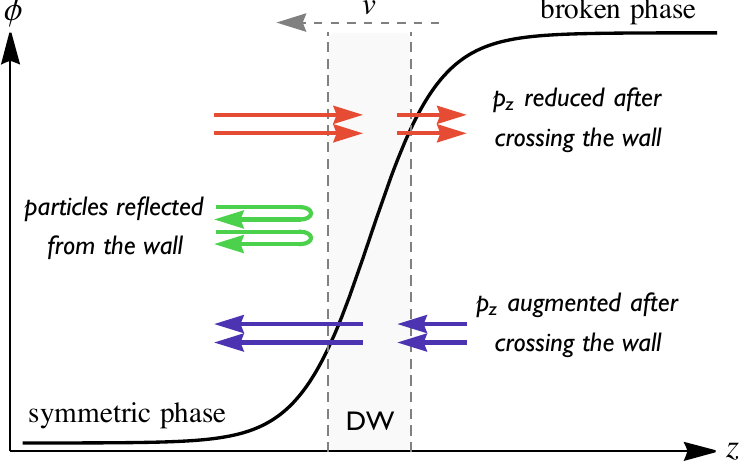}}
	\caption{{\it Left panel}: Paths in the collisionless limit with constant energy $E=\sqrt{p_\bot^2+p_z^2+m^2(z)}$ and transverse momentum in the $z - p_z$ phase space for $m(z)\propto 1 + \tanh(z/L)$. The red, green and blue curves denote three different classes of paths, as explained in the main text. The arrows show the direction of propagation the phase space. {\it Right panel}: Schematic representation of the particle's behaviour  across the domain wall (DW).}\label{fig:flow_paths}
\end{figure}

\subsection{Flow paths and the collision integral}\label{sec:Liouville}

Along paths where both the transverse momentum $p_\bot$ and the combination $p_z^2 + m^2(z)$ are constant, the Liouville operator reduces to a total derivative,
\begin{equation}
{\cal L} = \left(\frac{p_z}{E} \partial_z - \frac{(m^2(z))'}{2E} \partial_{p_z}\right) \quad \to \quad \frac{p_z}{E} \frac{d}{dz}\,.
\end{equation}
These paths are nothing but the trajectories in the $(p_\bot, p_z, z)$ phase space in the collisionless limit ${\cal C} \to 0$, along which both the energy and the momentum parallel to the wall are conserved. Three classes of solutions are found for ${\cal C}=0$. They are schematically shown in fig.~\ref{fig:flow_paths}, where, to simulate the case of a particle whose mass is generated by the field's value inside the bubble, $m(z)$ was modeled as $m(z)=m_0(1+\tanh(z/L))$, with $L$ the wall thickness.
Red curves describe particles that travel towards the bubble and have enough momentum to cross the wall, $p_z(-\infty)>m_0$. After crossing it, they lose part of their momentum in favour of the mass $m(z)$. 
Green curves describe particles that travel towards the wall with momentum $p_z(-\infty)<m_0$ and get reflected from it. Finally, blue curves describe particles that travel in the opposite direction and exit from the bubble, gaining momentum as the mass decreases to $0$ ($p_z(-\infty)$ will thus be $p_z(-\infty)<-m_0$).

\begin{figure}
	\centering
	{\includegraphics[width=.47\textwidth]{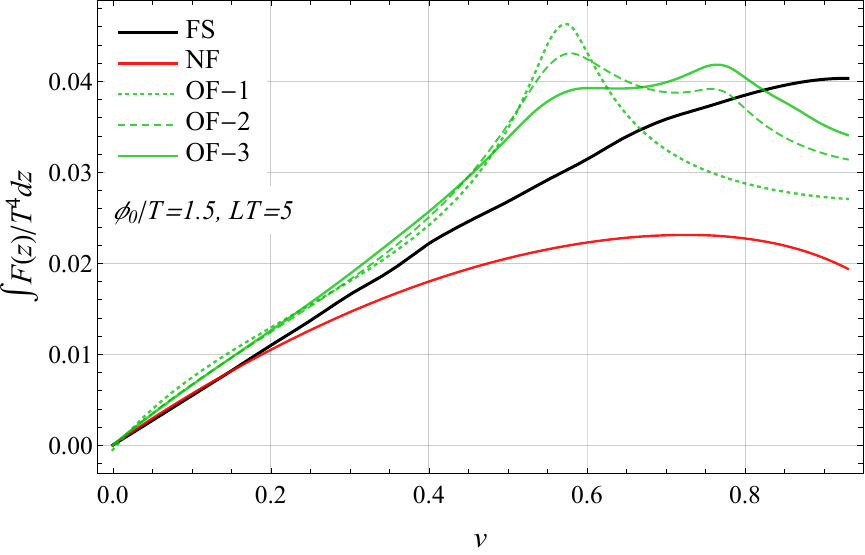}}
	\caption{Friction acting on the bubble wall as a function of the velocity $v$. The black curve shows the solution to the full Boltzmann equation (FS), the green curves the results obtained with the old formalism at order $1$, $2$ and $3$, respectively, and the red curve the result obtained with the new formalism~\cite{Laurent:2020gpg}.}\label{fig:friction}
\end{figure}

The collision integral for $2 \to 2$ processes is
\begin{equation}
{\cal C}[f] = \sum_i \frac{1}{4N_p E_p} \int\! \frac{d^3{\bf k}\,d^3{\bf p'}\,d^3{\bf k'}}{(2\pi)^5 2 E_k 2E_{p'} 2E_{k'}} |{\cal M}_i|^2
\delta^4(p+k-p'-k') {\cal P}[f]\,,
\label{collintegral}
\end{equation}
where $\mathcal P[f]$ is the population factor (the $\pm$ signs are for bosons and fermions, respectively)
\begin{equation}
{\cal P}[f] = f(p) f(k)(1 \pm f(p'))(1 \pm f(k')) - f(p') f(k')(1 \pm f(p))(1 \pm f(k))\,,
\end{equation}
$N_p$ are the degrees of freedom of the incoming particle with momentum $p$ whose distribution function we aim to calculate, $k$ is the momentum of the other incoming particle, and $p',\,k'$ are the momenta of the outgoing particles.
The sum runs over all $2\to2$ scattering processes with squared amplitude $|\mathcal M_i|^2$. As mentioned in the introduction, the expression \eqref{collintegral} for the collision integral can be greatly simplified by exploiting its symmetries to first write it as an Hermitian operator, and to express its eigenfunctions in spherical harmonics. This allows to swap the nine-dimensional integral for a matrix multiplication and to use a multipole expansion to identify the eigenfunctions that truly contribute to the collision integral. For details, we refer the reader to~\cite{DeCurtis:2023hil,DeCurtis:2024hvh}. In passing, let us also note that in \cite{DeCurtis:2024hvh} a strategy to go beyond the linear approximation \eqref{eq:Boltz_lin} and deal with non-linear terms was presented and implemented. We will refrain from discussing non-linear terms in this manuscript.

\section{Numerical analysis}\label{sec:numerical}

Writing as before $f=f_v+\delta f$, the collision integral can be split in two parts: one proportional to $\delta f(p)$, and another one where $\delta f$ depends on the integrated momenta $k, \,p', \,k'$. The linearized Boltzmann equation then takes the form 
\begin{equation}
\frac{d}{dz}\delta f-\frac{\mathcal Q(p)}{p_z}\frac{f_v}{f'v}\delta f= \mathcal S+\frac{f_v}{pz} \braket{\delta f}
\end{equation} 
where $\mathcal S$ is the $\delta f$-independent source term (see \eqref{eq:Boltz_lin}), and $\braket{\delta f}$ collectively denotes all the contributions where $\delta f$ is integrated.

\begin{table}[t]
	\centering
	\begin{tabular}{c|c|c|c||c|c|c|c}
		& $m_s\,$(GeV) & $\lambda_{hs}$ & $\lambda_s$ & $T_n\,$(GeV) & $T_c\,$(GeV) & $T_+\,$(GeV) & $T_-\,$(GeV)   \\
		\hline
		\rule{0pt}{1.1em}BP1 & 103.8 & 0.72 & 1 & 129.9 & 132.5 & 130.1 & 129.9 \\ 
		\rule{0pt}{1.em}BP2 & 80.0 & 0.76 & 1 & 95.5 & 102.8 & 96.7 & 95.5
	\end{tabular}
	\\
	\vspace{0.5cm}
	\begin{tabular}{c|c|c|c|c}
		& $v_w$ & $\delta_s$ & $L_h T_n$ & $L_s T_n$ \\
		\hline
		\rule{0pt}{1.1em}BP1 & 0.28\ \ [0.39]\ \ (0.57) & 0.78\ \ [0.79]\ \ (0.75) & 9.2\ \ [9.7]\ \ (8.1) &7.4\ \ [7.7]\ \ (6.7)\\ 
		\rule{0pt}{1.em}BP2 & 0.41\ \ [0.47]\ \ (0.61) &0.81\ \ [0.81]\ \ (0.81) & 5.1\ \ [5.2]\ \ (4.7) &4.2\ \ [4.3]\ \ (4.1)
	\end{tabular}
	\caption{{\it First row:} $m_s$, $\lambda_{hs}$ and $\lambda_s$ individuate the benchmark points BP1 and BP2. $T_n$ and $T_c$ are the nucleation and critical temperatures, $T_+$ and $T_-$ the temperatures in front and behind the DW. {\it Second row}: terminal values of the parameters $v_w$, $\delta_s$, $L_h$, $L_s$ for the two benchmark points.
		In each column, the first number is the result found including both the top quark and EW gauge bosons out-of-equilibrium perturbations. Numbers in square brackets are obtained neglecting the gauge bosons contribution, while the numbers in parentheses are obtained neglecting out-of-equilibrium perturbations altogether.}
	\label{tab:parameters_resultsW}
\end{table}

To solve the Boltzmann equation, a numerical iterative method was developed in~\cite{DeCurtis:2022hlx}, where $\braket{\delta f}$ terms are first ignored and, starting from $n=1$, the solution at iteration $n+1$ is obtained inserting the solution for $\delta f$ found at iteration $n$ in $\braket{\delta f}$. The strategy was first tested considering a single species in the plasma, the top quark, the state with the largest coupling to the Higgs, and taking the tanh ansatz for the Higgs field
\begin{equation}
	\label{tanh}
h(z)=\frac{h_0}{2}\left(1+\tanh\frac{z}{L_h}\right),
\end{equation}
where $L_h$ is the thickness of the bubble wall and  $h_0$ is the Higgs VEV in the broken phase. The numerical analysis in~\cite{DeCurtis:2022hlx} was carried out fixing the transition temperature $T = 100\;{\rm GeV}$, $L_h = 5/T$ and $h_0 = 150\;{\rm GeV}$.

The out-of-equilibrium friction acting on the bubble wall ($N$ are the degrees of freedom), 
\begin{equation}
F(z) = \frac{d m^2}{d z} N \int \frac{d^3 {\bf p}}{(2 \pi)^3 2E} \delta f(p)\,,
\end{equation}
was numerically evaluated. The result is reported in Fig.~\ref{fig:friction}, where the integrated friction thus found is shown as a function of the wall velocity  (black curve) together with previous results (green and red curves). In particular, green curves show the integrated friction obtained in the old formalism~\cite{Moore:1995si} when the fluid approximation is improved with higher-order corrections ($1$, $2$ and $3$)~\cite{Dorsch:2021ubz}. The red curve shows the result obtained in the new formalism~\cite{Laurent:2020gpg}. 
A fair numerical agreement is found between all the solutions for $v_w \lesssim 0.3$. At higher velocities, the OF's solution develops peaks that correspond to zero eigenvalues of the Liouville operator. The result for the full solution confirms that the latter are an artifact of the weight strategy. The NF's solution does not show these peaks but turns out to be in agreement with the full solution only for small velocities $v_w\lesssim 0.3$.

\begin{figure}
\includegraphics[scale=0.7]{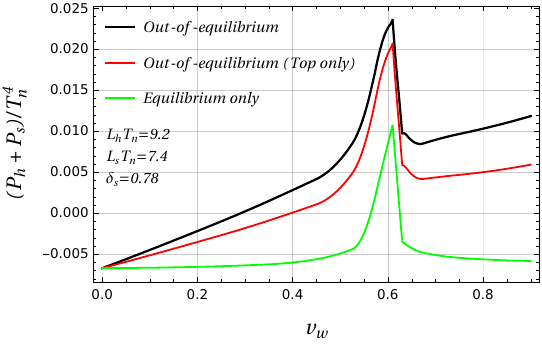}
\includegraphics[scale=0.7]{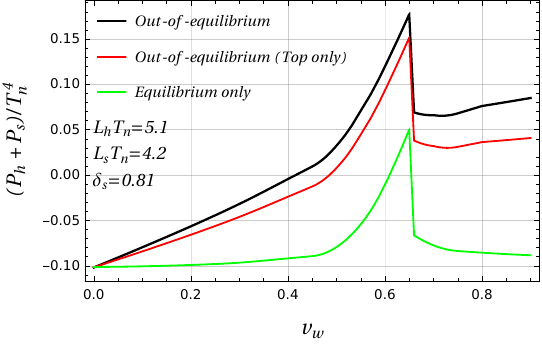}
\caption{Total pressure $P_h+P_s$ as a function of the wall velocity $v_w$ 
	for the two benchmark points BP1 ({\it left panel}) and BP2 ({\it right panel}) in Table~\ref{tab:parameters_resultsW}. The green curves are obtained assuming all species are at equilibrium, the red ones including out-of-equilibrium effects from the top quark, and the black ones including out-of-equilibrium effects from the top quark and EW gauge bosons.}
\label{fig:pressure}
\end{figure}

The bubble wall velocity was later calculated in the SSM (for two benchmark points) including the top quark perturbations only~\cite{DeCurtis:2023hil}, and adding the EW gauge boson contributions~\cite{DeCurtis:2024hvh}. Two-step phase transitions where the EW vacuum is reached from an intermediate phase with non-vanishing singlet were analysed, and the tanh ansatz was taken for both the Higgs and the singlet ($s(z)=s_0\left(1+\tanh(z/L_s+\delta_s)\right)$ supplements Eq.\,\eqref{tanh}). Hydrodynamic equations are used together with the scalar equations of motion (EOM) to determine $v_w$, the tilt $\delta_s$ in the scalar profiles and the wall thicknesses $L_h$ and $L_s$ as the values that make the pressure exerted on the wall and its gradient vanish, allowing for a steady state solution. The pressure on the wall in the Higgs and in the singlet direction are denoted with $P_h$ and $P_s$, respectively. At zeroth order in the proposed iterative procedure, the calculation of the wall velocity through the combined solution of the hydrodynamic equations and of the scalar EOM coincides with the equilibrium determination of $v_w$. The numerical algorithm we developed to solve these equations already allows, at this stage, to calculate the wall velocity at equilibrium in the full parameter space of BSM theories in a fair amount of time~\cite{in preparation}.

The results for $v_w$ are summarized in Fig.\,\ref{fig:pressure}, where we plot the total pressure in terms of $v_w$, and in Table \ref{tab:parameters_resultsW}, where the chosen benchmark points are individuated through the values of the singlet mass $m_s$ in the EW vacuum, of the portal coupling $\lambda_{hs}$, and of the singlet's self-coupling $\lambda_s$. The impact of out-of-equilibrium contributions from both the top and the EW gauge bosons can be appreciated.

\section{Conclusions and Outlook}\label{sec:conclusions}

In~\cite{DeCurtis:2022hlx} it was presented for the first time a numerical solution to the full Boltzmann equation describing the distribution function of particles during a first order phase transition. Differently from other approaches in the literature, we did not rely on any ansatz for the out-of-equilibrium distribution functions. The results were critically compared with those found using the traditional strategies put forward in the literature, namely the fluid approximation~\cite{Moore:1995si}, its extended version~\cite{Dorsch:2021ubz,Dorsch:2021nje}, and the new formalism~\cite{Laurent:2020gpg}. The efficiency of the algorithm developed in \cite{DeCurtis:2022hlx} was later improved in \cite{DeCurtis:2023hil} and \cite{DeCurtis:2024hvh}, where we also calculated the wall velocity for benchmark points in the singlet extension of the Standard Model. 

These results represent a great advancement toward a reliable characterization of the bubble wall dynamics. The improved computational performances obtained with the refinements in \cite{DeCurtis:2023hil} and \cite{DeCurtis:2024hvh} of the algorithm proposed in \cite{DeCurtis:2022hlx} open the door to the possibility of scanning the parameter space of BSM models and studying experimentally falsifiable signals (gravitational waves, baryon asymmetry and so on) therein. This ambitious program will be inaugurated by the equilibrium calculation of $v_w$ in several BSM models of phenomenological interest \cite{in preparation}.


\begin{thebibliography}{45}

\bibitem{Caprini:2015zlo}
C.~Caprini et~al., JCAP \textbf{04}, 001 (2016), \texttt{1512.06239}

\bibitem{Caprini:2019egz}
C.~Caprini et~al., JCAP \textbf{03}, 024 (2020), \texttt{1910.13125}

\bibitem{Kawamura:2006up}
S.~Kawamura et~al., Class. Quant. Grav. \textbf{23}, S125 (2006)

\bibitem{Kawamura:2011zz}
S.~Kawamura et~al., Class. Quant. Grav. \textbf{28}, 094011 (2011)

\bibitem{Hu:2017mde}
W.R. Hu, Y.L. Wu, Natl. Sci. Rev. \textbf{4}, 685 (2017)

\bibitem{Ruan:2018tsw}
W.H. Ruan, Z.K. Guo, R.G. Cai, Y.Z. Zhang, Int. J. Mod. Phys. A \textbf{35},
  2050075 (2020), \texttt{1807.09495}

\bibitem{TianQin:2015yph}
J.~Luo et~al. (TianQin), Class. Quant. Grav. \textbf{33}, 035010 (2016),
  \texttt{1512.02076}

\bibitem{Coleman:1977py}
S.~R.~Coleman,
Phys. Rev. D \textbf{15}, 2929-2936 (1977)
[erratum: Phys. Rev. D \textbf{16}, 1248 (1977)]

\bibitem{Callan:1977pt}
C.~G.~Callan, Jr. and S.~R.~Coleman,
Phys. Rev. D \textbf{16}, 1762-1768 (1977)

\bibitem{Linde:1977mm}
A.~D.~Linde,
Phys. Lett. B \textbf{70}, 306-308 (1977)

\bibitem{Ignatius:1993qn}
J.~Ignatius, K.~Kajantie, H.~Kurki-Suonio and M.~Laine,
Phys. Rev. D \textbf{49}, 3854-3868 (1994) \texttt{ astro-ph/9309059}


\bibitem{Laurent:2022jrs}
B.~Laurent, J.~M.~Cline,
Phys. Rev. D \textbf{106}, 023501 (2022),
\texttt{2204.13120}



\bibitem{DeCurtis:2023hil}
S.~De Curtis, L.~Delle Rose, A.~Guiggiani, \'A.~Gil Muyor, G.~Panico,
JHEP \textbf{05}, 194 (2023)
\texttt{2303.05846}.

\bibitem{DeCurtis:2024hvh}
S.~De Curtis, L.~Delle Rose, A.~Guiggiani, \'A.~Gil Muyor, G.~Panico,
JHEP \textbf{05}, 009 (2024), \texttt{2401.13522}.

\bibitem{DeCurtis:2022hlx}
S.~De~Curtis, L.D. Rose, A.~Guiggiani, A.G. Muyor, G.~Panico, JHEP \textbf{03},
163 (2022), \texttt{2201.08220}

\bibitem{in preparation}
C.~Branchina, et al., {\it in preparation}


\bibitem{Moore:1995ua}
G.D. Moore, T.~Prokopec, Phys. Rev. Lett. \textbf{75}, 777 (1995),
  \texttt{hep-ph/9503296}

\bibitem{Moore:1995si}
G.D. Moore, T.~Prokopec, Phys. Rev. D \textbf{52}, 7182 (1995),
  \texttt{hep-ph/9506475}

\bibitem{Cline:2020jre}
J.M. Cline, K.~Kainulainen, Phys. Rev. D \textbf{101}, 063525 (2020),
  \texttt{2001.00568}

\bibitem{Laurent:2020gpg}
B.~Laurent, J.M. Cline, Phys. Rev. D \textbf{102}, 063516 (2020),
  \texttt{2007.10935}

\bibitem{Dorsch:2021ubz}
G.C. Dorsch, S.J. Huber, T.~Konstandin, JCAP \textbf{08}, 020 (2021),
  \texttt{2106.06547}

\bibitem{Dorsch:2021nje}
G.C. Dorsch, S.J. Huber, T.~Konstandin (2021), \texttt{2112.12548}



\end{thebibliography}
\end{document}